\documentstyle[preprint,aps,prl]{revtex}
\newcommand{\be}{\begin{equation}} \newcommand{\ee}{\end{equation}}

\newcommand{\bea}{\begin{eqnarray}} \newcommand{\eea}{\end{eqnarray}}
 \begin{document} \vskip 3.5cm

\title{Source lifetime dependence of
neutrino oscillations - a simple wave-function derivation}

\author{Subhendra Mohanty}

\address{{\it Theory Group, Physical Research Laboratory, \\
Ahmedabad - 380 009, India }}

\maketitle

\begin{abstract}

{When neutrinos are produced from long lived particles like pions,
kaons and muons, the life-time of the source particle affects the
flavour conversion probability formula. 
Experiments like LSND which use muon decay neutrinos are two orders of
magnitude more sensitive to lower values of mass square difference
compared to other experiments where the sources are pions or kaons.
}

\end{abstract}
\bigskip
The standard formula for neutrino oscillations is derived with the
assumption \cite{wp} that the 
uncertainity in the initial position of the neutrino is small compared to 
the distance between the
production and the detection sites of the neutrinos . We shall show here that in a covariant
treatment, the effective spread of the netutrino wavefunction in space is
given by $(\sigma_x + c \tau)$ where $\sigma_x$ is the initial spatial
uncertainity of the production point of the neutrinos and $\tau$ is the
lifetime of the sources (eg. pions, kaons , muons etc) whose decay produce
the
neutrinos. We find that the effect of long lifetime sources is to
exponentially 
suppress the oscillation term in the conversion probability fromula.  The
flavor conversion probability as a function of distance , for relativistic
neutrinos produced from long lived resonances, turns out to be
\be  
P(\nu_{e} \rightarrow \nu_{\mu};X)={1\over2}~sin^2 2\theta ~
~~(1-~cos\{{2.53 \Delta m^2 X \over E}\}~~exp\{-({1.79 \Delta m^2 \tau
\over
E})^2\} ~~)
\label{cor1}
\ee
where $\Delta m^2$ is the mass square difference in $eV^2$ ,$L$ where
$\Delta m^2$ is the mass square difference in $eV^2$ ,$X$ is the
detector distance in $m~(km)$,  $\tau$ is the lifetime of the source
particle
in the lab frame
in $m~(km)$ and $E$ is the  energy
in $MeV~(GeV)$. When $c\tau$ is comparable to the detector distance
spatial oscillations of the conversions probability is not seen. However
the neutrino mass squared difference that can be probed  is lower
for longer lived sources. We find that for the LSND experiment where the
neutrinos are produced from muons ($c\tau_{\mu} =658.65 m$), when fitted
with
the covariant oscillation formula (\ref{cor1}) gives a bound on $\Delta
m^2$ which is two orders of magnitude lower than the corresponding bounds
obtained from other similar experiments but where the neutrino sources are
pions or kaons ($c\tau_{\pi}= 7.8 m$).

We start with an initial wave function of a general Gaussian form
\be
\Psi^{in}_a(x,t)= {1\over (2\pi \sigma_x
\sigma_t)^{1/2}}~exp\{~~i (P_a (x-x_i)-E_a(t-t_i) )~-
~{(x-x_i)^2\over4 \sigma_x^2}~-~{(t-t_i)^2\over 4\sigma_t^2}~\}
\label{psi-in}
\ee
The initial spread of the wavefunction in space around the mean initial
position $x_i$ is denoted by $\sigma_x$ and the spread in the time
direction around the mean initial time of production $t_i$ is denoted by 
$\sigma_t$. The magnitudes of $\sigma_x$ and $\sigma_t$ depends on how the
state is prepared. The time evolved wavefunction can be obtained from
the initial wavefunction (\ref{psi-in}) by the linear superposition
principle , which can be written formally as
\be
\Psi_a(x_f,t_f)= \int ~dx~dt~~K(x_f - x, t_f -t) ~~\Psi^{in}_a(x,t)
\label{psi-out}
\ee
where $K(x_f-x,t_f-t)$ is the probability amplitude of a particle
initially located at $(x,t)$ to be detected at another spacetime
point $(x_f,t_f)$.
The transition amplitude  $K(x_f-x,t_f-t)$  for a free particle is given
by the expression (\cite{pp,nupb,bogol})
\be
K(x_f-x ,t_f-t;m_a) = ({i\over 4 \pi^2}) ~({m_a
\over s})~ K_1(i m_a s) 
\label{three}
\ee
where  $s= ((t_f-t)^2 - ( x_f-x)^2 )^{1/2}$ is the
invariant spacetime interval propagated by the $\nu_i$ mass eigenstate. If
this interval is large ($s >> m_a ^{-1}$) and time-like ( $(t_f-t) \ge
(x_f-x)$ ) then we can use the asymptotic expansion of the Bessel
function 
\be
~K_1(i m s)
\simeq \left({2 \over \pi i m s} \right)^{1/2} ~exp\{-i m s\}
\label{bessel}
\ee   
to obtain from (\ref{three}) the expression for the propagation amplitude
at large time-like separation
\be
K(x_f-x,t_f-t; m_a) =({m_a\over 2\pi i
\sqrt{(t_f-t)^2 -( x_f-x)^2 }})^{1/2} ~ 
 exp\{ -im_a \sqrt{(t_f-t)^2
 -(x_f - x)^2} \}
\label{four}
\ee
This expression for the free particle propagator is valid for bosons. For
fermions their is a extra factor $(i \not \partial + m_a )$ operating on
the l.h.s of the expressions (\ref{three}) and (\ref{four}). This factor
only changes the normalisation and the expressions for the
conversion probability is
identical for bosons and fermions. The time evolved wave-function
can be evaluated by substituting the expression
(\ref{four}) for the
propagator $K(x_f-x,t_f-t)$ in the expression (\ref{psi-out}) for
$\Psi$, 
\bea
\Psi_a(x_f-x_i,t_f-t_i;m_a)&=&  N \int~dx~dt~
({1\over 
(t_f-t)^2 -( x_f-x)^2 })^{1/4} \nonumber\\[8pt]
~&\times& exp \{i \Phi(x,t) -
~{(x-x_i)^2\over4 \sigma_x^2}~-~{(t-t_i)^2\over 4\sigma_t^2}~\}
\label{psi-out2}
\eea
where the phase factor in the exponential is of the form
\be
\Phi(x,t)= -m_a \sqrt{(t_f-t)^2
 -(x_f - x)^2} + P_a (x-x_i)-E_a(t-t_i) 
\label{phi}
\ee
and the constant coefficients have been clubbed together as the factor 
$ N$
which normalises the wave-function. We perform the integrations in
(\ref{psi-out2}) by the method of stationary phases \cite{wolf}. The
integral is approximated by the integrand along the trajectory where the
phase is an extremum. The extremum of the phase (\ref{phi}) is given by
$ (\partial \Phi/ \partial t) =0 \Rightarrow (m_a ~(t_f -t)/  
 ((t_f-t)^2-(x_f - x)^2)^{1/2}) = E_a  $ and $
(\partial \Phi / \partial x) =0 \Rightarrow (m_a ~(x_f -x)/
 ((t_f-t)^2-(x_f - x)^2)^{1/2}) = P_a  $.
One can solve for $t$ and $x$ using these two equations 
 and substitute in the integrand of (\ref{psi-out2}) to get
the expression for the integral in the stationary phase approximation ,
which turns out to be
\be
\Psi_a(x_f-x_i,t_f-t_i)= N^{\prime} exp~\{-i E_a(t_f-t_i) +i 
P_a(x_f-x_i)-{((x_f-x_i)- (t_f-t_i){P_a\over E_a})^2\over 4(\sigma_x^2 +
\sigma_t^2 ({P_a\over E_a})^2)}\}
\label{psi-out3}
\ee
where again we have clubbed together all the constants as the
normalisation coefficient $N^{\prime}$. Denoting $X=x_f-x_i$, $T=t_f-t_i$
and $v_a= (P_a/E_a)$ which can be identifired by distance of propagation,
time of propagation and the group velocity of the particle respectively we
can  write the expression for the time evloved wave-function compactly as
\be
\Psi_a(X,T)= N^{\prime} exp~\{-i E_a T +i
P_a X-{(X- v_a T )^2\over 4(\sigma_x^2 +
v_a^2 \sigma_t^2 )} \}
\label{psi-out4}
\ee
The state vector of a mass eigenstate can be in the mass basis $|\nu_a>$
in the form
\be
|m_a;X,T> = \Psi_a(X,T) |\nu_a>
\ee
On the other hand the state of a weak interaction state say $\nu_e$ is
expressed as a linear combination
\be
|\nu_e;X,T> = \sum_a  \Psi_a(X,T) |\nu_a><\nu_a|\nu_e>
\label{nue}
\ee
The probability amplitute of a neutrino to be produced at $(x_i,t_i)$ as
a $\nu_e$ and to be detetected at $(x_f,t_f)$ as another weak eigenstate
$\nu_\mu$ is  given by
\be
{\cal A}(\nu_e \rightarrow \nu_\mu ; X,T)=<\nu_\mu|\nu_e;X,T>=
\sum_a  \Psi_a(X,T)~<\nu_\mu |\nu_a><\nu_a|\nu_e>
\label{amp}
\ee
and the corresponding probability is given by
\bea
P(X,T;\nu_e \rightarrow \nu_\mu)&=&|<\nu_\mu|\nu_e;X,T>|^2 
\nonumber\\[8pt]
&=&\sum_{a,b}  \Psi_a(X,T)\Psi_b^*(X,T)
~<\nu_\mu|\nu_a><\nu_a|\nu_e>~~<\nu_e|\nu_b><\nu_b|\nu_\mu>
\label{P}
\eea
Restricting ourselves to mixing between two flavour generations, and
denoting the mixing matrix elments as
$<\nu_a|\nu_e>=-<\nu_b|\nu_\mu>=sin\theta$ and 
$<\nu_a|\nu_\mu>=<\nu_b|\nu_e>=cos\theta$, the expression (\ref{P}) for
the probability reduces to the form
\bea
P(\nu_e \rightarrow \nu_\mu ;X,T)&=&sin^2 \theta~ cos^2 \theta~\{ 
\Psi_a(X,T)\Psi_a^*(X,T)+
\Psi_b(X,T)\Psi_b^*(X,T)\} \nonumber\\[8pt]
&-& sin^2\theta ~cos^2\theta ~\{\Psi_a(X,T)\Psi_b^*(X,T) +
\Psi_a^*(X,T)\Psi_b(X,T)\}
\label{P1}
\eea
This expression expresses the probability of a $\nu_e$ produced at
$(x_i,t_i)$ to be converted to  a $\nu_\mu$ at some other spactime 
point $(x_f,t_f)$. In actual experiments only the distance $X$ between the
source and the detector is known but the arrival time of the neutrinos is
not measured. The expression for the conversion probability as a function
of only $X$ is obtained by taking the time average of (\ref{P1}).
The normalisation is chosen so that $\Psi_a(X,T)$ represents a single
particle wave-function and therefore,
\be
\int_{-\infty}^{+\infty} dT \Psi_a(X,T)\Psi_a^*(X,T) = 1
\ee
The non-trivial contribution to the time averaging comes from the
interference term of   (\ref{P1})
\bea
Re ~\int_{-\infty}^{+\infty} dT \Psi_a(X,T)\Psi_b^*(X,T) &=& Re
~\int_{-\infty}^{+\infty} dT ~ exp~\{-i (E_a -E_b)T +i (P_a -P_b) X\}
\nonumber\\[8pt]
&\times& exp\{-{(X-v_a T )^2\over 4(\sigma_x^2 +
v_a^2 \sigma_t^2 )} -{(X-v_b T )^2\over 4(\sigma_x^2 +
v_b^2 \sigma_t^2 )} \}
\label{int}
\eea
This integral can be evaluated by the completing the squares, to give
\bea
Re ~\int_{-\infty}^{+\infty} dT \Psi_a(X,T)\Psi_b^*(X,T) &=& 
cos\{ ((E_a -E_b){(v_a +v_b)\over (v_a^2 +v_b^2)} - (P_a
-P_b))X \}
\nonumber\\[8pt]
&\times& exp\{-{1\over (v_a^2 +v_b^2) }((E_a-E_b )^2 {\bar \sigma}^2
-{(v_a-v_b )^2 X^2 \over 4 {\bar \sigma}^2 })  \}
\label{int1}
\eea
where 
\be
 {\bar \sigma} \equiv (\sigma_x + {(v_a^2 + v_b^2)\over (v_a + v_b) }
\sigma_t).
\ee
We express the energy, momentum and the masses  in terms
of
their averages $E=(E_a +E_b)/2, P=(P_a +P_b)/2, m=(m_a+m_b)/2$ and
differences $\Delta E=(E_a-E_b),\Delta P=(P_a-P_b), \Delta m= (m_a -
m_b)$.
 In terms of these variables (\ref{int1}) reduces to the form 
\bea
{\sqrt 2\over (v_a^2 + v_b^2)^{1/2}}~cos~(~(E \Delta E- P \Delta P){X\over
P} )~~e^{-A}~~~~~~,\\[8pt]
\label{int2}
A~=~ ~(\Delta E)^2 ~ {{\bar \sigma}^2\over 2}
({E^2\over P^2}) ~+~({\Delta P^2
\over P})^2 {X^2\over 4{\bar \sigma}^2} 
\label{A}
\eea
where we have retained the terms to the first order in $\Delta E/E, \Delta
P/P $ and $\Delta m/m$.
  The  energy and
momenta of the neutrinos are determined by the energy-mommentum
conservation laws at the
production vertex. Since niether the energy nor the momenta of the
remaining outgoing particles are measured, one cannot fix either the
energy or the momenta of the different neutrino mass eigenstates
in the linear combination state. So one cannot assume either
$E_a =E_b$ or $P_a=P_b$. Only the mass shell relation
 $E_a^2 = P_a^2 + m_a^2 $ for each mass eigenstate
$\nu_a$ (\cite{pp,nupb}). In terms of the average and difference , the
mass shell conditions imply,
\be
E \Delta E - P \Delta P = {\Delta m^2 \over 2}
\label{ms}
\ee
Using the relation (\ref{ms}) in (\ref{int2}) we see that the interference
term reduces to the form
\be
{\sqrt 2\over (v_a^2 + v_b^2)^{1/2}}~cos~(~{\Delta m^2\over
2 P}X )~~e^{-A}
\label{int3}
\ee
we see that the oscillation length is $L_{osc} = (4 \pi P/\Delta m^2)$ 
for both relativistic as well as non-relativistic particles.
The significant difference in the formula comes from the suppression
factor $A$ (\ref{A}).
For intereference term to be non-zero, both the terms of the suppression
factor must be small. The second term in (\ref{A})
is $<< 1$ as long as
$X << L_{coh}$ with  the coherence length for relativistic neutrinos 
given by ,
\be
L_{coh} = 4 {\sqrt 2} (\sigma_x + v \sigma_t)~({ E^2 \over \Delta m^2})
\label{xcoh}
\ee
where $v=(v_a^2 + v_b^2)/(v_a + v_b) \simeq 1 $.
This expression for the coherence length differs from the expression
derived in the standard wave-packet treatments {\cite{wp}} by the
presence of the $\sigma_t$ term. In all accelerator neutrino oscillation 
experiments the
criterion $X << L_{coh}$ is satisfied , so the effect of this term is 
negligible. 
The first term of (\ref{A}) plays a more significant role. For
relativistic neutrinos, the suppresion factor (\ref{A}) reduces to the
form,
\be
A~=~ ({\Delta m^2 ~ (\sigma_x + v \sigma_t)\over 2 {\sqrt 2} E})^2
\label{A1}
\ee
In the accelerator experiments $\sigma_x$ - the spread in the beam
of
primary particles is of the order of a few $cm$. The neutrinos are
produced from the secondary decays of pions and kaons produced in the
primary collision. The uncertainity in time of production of such
neutrinos $\sigma_t$ is then given by the lifetime $\tau$ of the 
pions, kaons or muons
which are the source of neutrinos for that particular experiment. In such
situations $\sigma_x << \sigma_t$ and $(\sigma_x +v \sigma_t) \simeq ~v
\tau$ and the expression
(\ref{A1}) reduces to
\be
 A~\simeq~ ({\Delta m^2 ~
\tau\over 2
{\sqrt 2} E})^2 .
\ee
 
Using these results we see that the expression for the time average of the
conversion probabilty, for relativistic neutrinos produced from long lived
resonances is given by  
\be
P(\nu_{\mu} \rightarrow \nu_{e};X)={1\over2}~sin^2 2\theta ~
~~(1-~cos({\Delta m^2 X \over 2 E})~~exp-({\Delta m^2 \tau \over
2 {\sqrt 2} E})^2 ~~)
\label{cor}
\ee
where $\Delta m^2$ is the mass square difference ,$X$ is the
detector distance ,  $\tau$ is the lifetime of the source
particle
in the lab frame
and $E$ is the neutrino energy. In practice one averages over the energy
flux $n(E)$ of the neutrinos and the average probability which is fitted
with the experimental number to obtain the allowed regions of $\Delta m^2$
and $sin^2 2 \theta$ is given formally by the expression,
\be
\langle P(\nu_{\mu} \rightarrow \nu_{e};X) \rangle = {\int~ dE~n(E) 
P(\nu_{\mu} \rightarrow \nu_{e};X) \over \int~ dE~n(E)}
\label{corE} 
\ee

 The limits on the values of $\Delta m^2$ and $sin^2
2\theta$ obtained by fitting the results of different experiments with
the covariant oscillation formula (\ref{cor}) are listed in Table I and
plotted in Fig 1, Fig 2 and Fig 3.
The regions
of the parameter space allowed by the covariant wavepacket formula and by
the standard formula  of the LSND (muon source) experiment are shown in
Fig.1.
and the Karmen (pion source) experiment are shown in Fig 2. Since the pion
decay length is of
the same order as the experimental baseline, the improvement in
sensitivity
to $\Delta m^2$ is marginal compared to LSND.  The combined result of all
experiments listed in Table I plotted using the covariant oscillation
formula is shown in Fig 3.

The dependence of the oscillation term on the source lifetime can
be understood as follows. When the time uncertainity is large the
uncertainity in energy becomes small and the wave-functions are
then eigenstates of energy. In the time smearing the overlap of two
different energy eigenstates dissapears and therefore the oscillation term
vanishes. 

{\it Acknowledgments} I thank Terry Goldman,
Walter Grimus,
 Harry Lipkin and Eligio Lisi for their comments on the earlier papers
\cite{pp,nupb}.

\begin{table}\squeezetable
\caption{The asymptotic limits on $\Delta m^2$ and $sin^2 2 \theta$
from different experments according to the
oscillation formula (\ref{cor}). $\tau$ is the lifetime of the neutrino
source in the lab frame, $<E_{\nu}>$ is the average $\nu$ energysource in
the lab frame, $<E_{\nu}>$ is the average $\nu$ energy, $X$ is
the detector
distance and $P$ is the experimental value of the
conversion
probability.}

\label{Sig} \begin{tabular}{|c|c|c|c|c|c|c|}
 Experiment(Source) &$\tau~(m) $  &$<E_{\nu}>~(MeV)$& $X~(m)$   
&  $P$ & $\Delta m^2~(eV^2)$ &$ sin^2
2 \theta$ \\
\tableline
LSND ($\mu$)  \cite{LSND}& $658.6$  &$ 30 $&$ 30 $& $(0.16 -
0.47)\times 10^{-2}$& $(1.0-1.5 )\times 10^{-3}$& $0.003 - 0.009$ \\  
\tableline
LSND ($\pi$)\cite{LSNDpi}& $17$&$130$&$30$&  $(0.26 \pm
0.15)\times 10^{-2}$&
$0.4-0.8 $& $0.002-0.0082$\\
\tableline
Karmen ($\pi$)\cite{karm}& $7.8$&$29.8$&$17.5$&  $< 0.3
\times
10^{-2}$& 
$< 0.08$& $< 0.6 \times 10^{-2}$\\
\tableline
E776 ($\pi$) \cite{E776}& $578$&$5\times10^3$&$10^3$& 
 $< 0.15\times 10^{-2}$&$< 0.1$& $< 0.3\times 10^{-2}$\\
\tableline
CCFR( $K$)\cite{CCFR}& $5.41 \times 10^3$ & $140 \times 10^3$ & $1.4
\times
10^3$&
$<0.9 \times 10^{-3}$& $<0.6$ & $< 0.18 \times 10^{-3}$\\
\tableline
Bugey(U,Pu) \cite{BUGEY}& $3 \times 10^{10}$ & $5$& $95$ &
 $ < 0.75 \times 10^{-1}$ & $< 10^{-9} $ & $<0.15$\\
\tableline
\end{tabular}
\end{table}

\begin{figure}
\vskip 20cm
\includegraphics{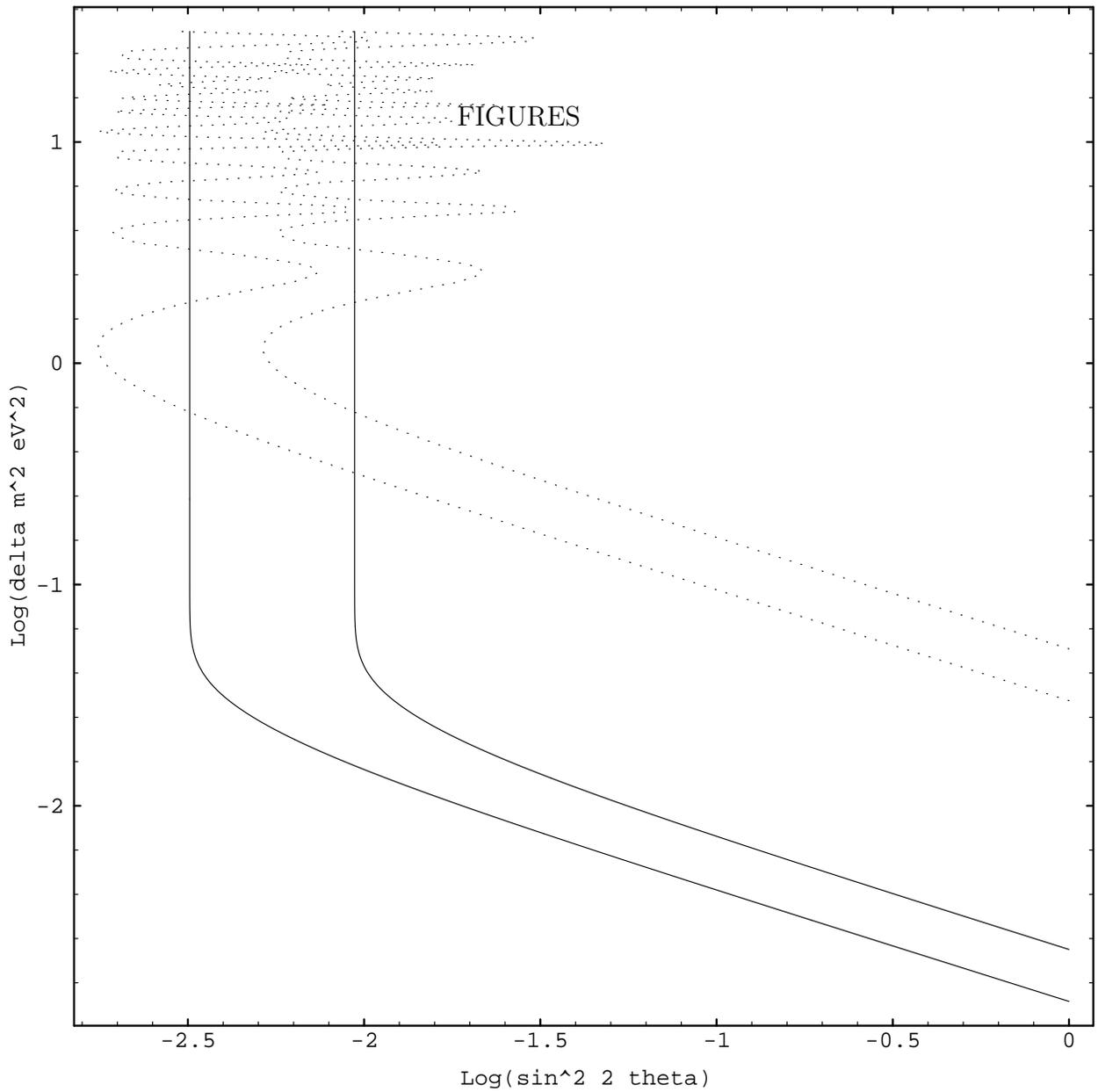}
\caption[dummy]{ Lsnd $\mu$ decay at rest experiment \cite{LSND} allowed
regions with
the standard formula
(between the dotted
curves)and the covariant
oscillation formula (continuous curves).
}

\label{fg:lsnd}
\end{figure}
\vskip 0cm
\newpage
.
\begin{figure}
\vskip 20cm
\includegraphics{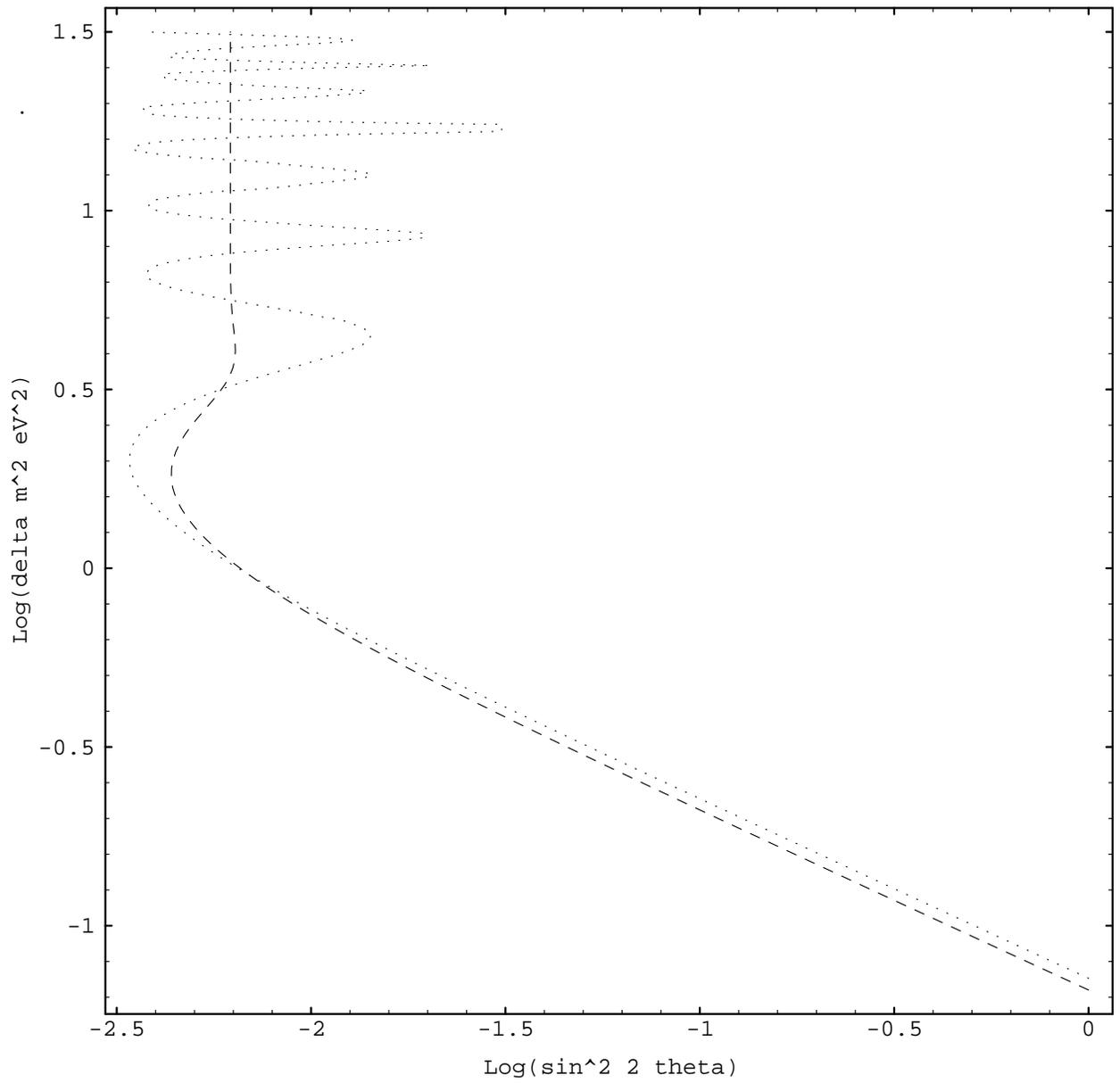}
\caption[dummy]{ Karmen $\pi$ decay at rest experiment \cite{karm} allowed
regions with the standard formula
(dotted line) and the
covariant oscillation formulas (dashed curve) .
}
\label{fg:karm}
\end{figure}
\vskip 0cm
\newpage
.
\begin{figure}
\vskip 18cm   
\includegraphics{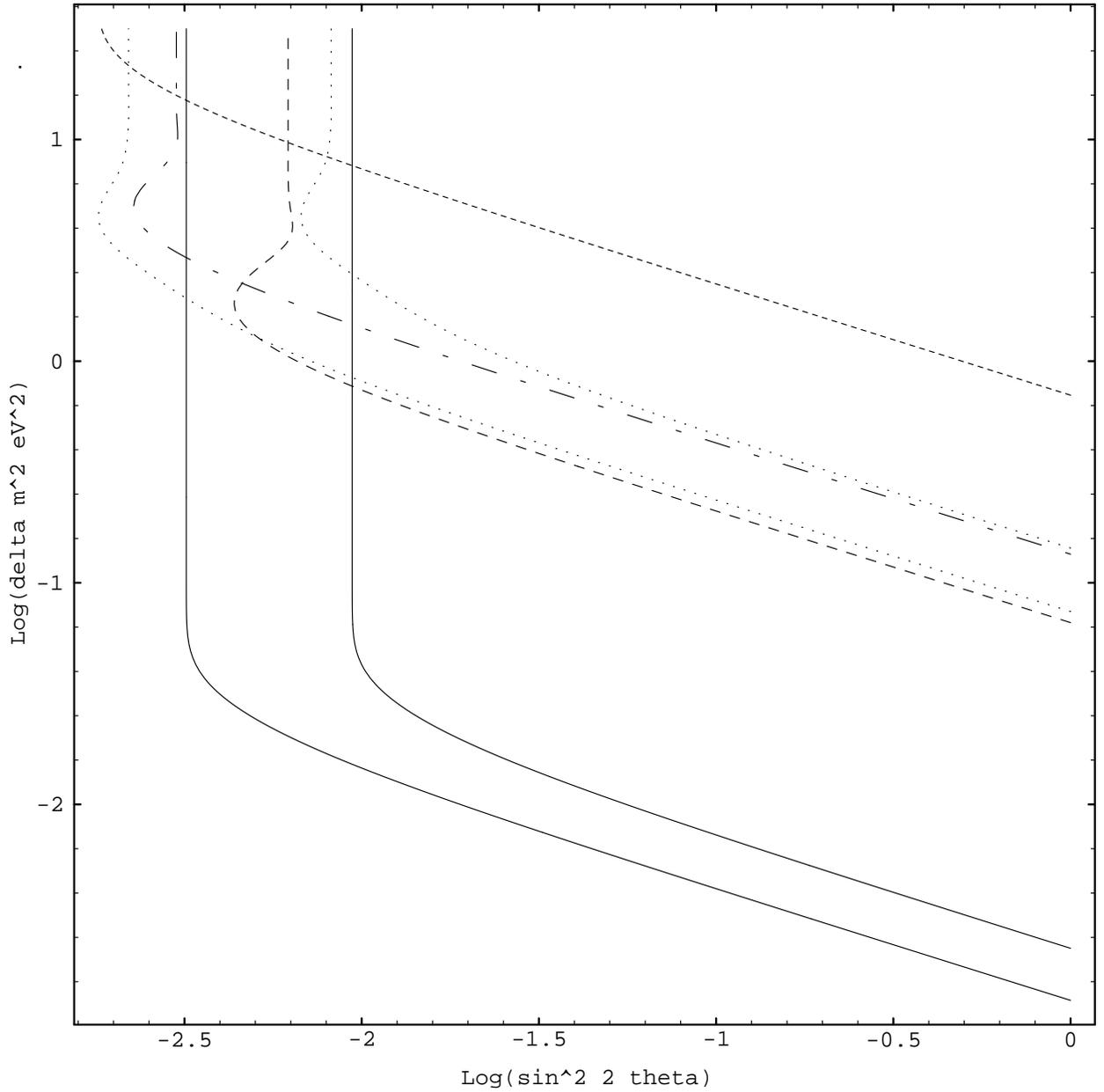}
\caption[dummy]{The region between the continuous lines is allowed by
the LSND $\mu$
experiment \cite{LSND} .The region between the dotted
lines is
allowed by the
LSND $\pi$ experiment \cite{LSNDpi} . Region ruled out
by E776 \cite{E776} is above the dashed-dotted curve and by Karmen
\cite{karm} is above dashed
curve. The region above the top-most dashed curved is ruled out by
CCFR \cite{CCFR}.}

\end{figure}


\vskip 0cm




\begin{references}

\bibitem{wp}C. W. Kim and A. Pevsner, {\it Neutrinos in
              Physics and Astrophysics} ,harwood academic publishers,
              Chur,(1993) Pg 206;\\
              C. Giunti, C. W. Kim, J. A. Lee and U. W. Lee,
             Phys. Rev. D {\bf 48}, 4310 (1993);\\ 
              B. Kayser, Phys. Rev. D {\bf 24}, 110 (1981);\\
              S.Nussinov, Phys Lett{\bf B63} , 201 (1976).\\
\bibitem{pp} S.Mohanty , "Production process dependence of
neutrino flavor conversions"  hep-ph 9706328.
\bibitem{nupb} S.Mohanty, "Covariant treatment of flavor oscillations" 
hep-ph  9702424.
\bibitem{bogol} N.N. Bogoliubov and D.V. Shirkov, {\it An introduction to
the theory of quantised fields }, Wiley Interscience, New
              York, (1959).
\bibitem{wolf} L. Mandel and E.Wolf, {\it Optical coherence and quantum
optics},
p 128, Cambridge University Press, (1995). 
\bibitem{LSND}  C.Athanassopoulos et al (LSND ) , Phys Rev Lett. {\bf 75}
, 2650 (1995).
\bibitem{E776} L.Brodovsky et al (BNL-E776), Phys. Rev Lett {\bf 68}, 274
(1992).
\bibitem{karm} B.Armbrusuter et al (Karmen), Nucl.Phys. {\bf B38 },235
(1995).
\bibitem{CCFR} A. Romosan et al (CCFR), Phys. Rev. Letters  ,{\bf 78},
2912(1994).
\bibitem{BUGEY} B.Achkar et al (Bugey), Nucl. Phys. {\bf B434} , 503
(1995).  
\bibitem{LSNDpi} C.Athanassopoulos et al (LSND ) , nucl-ex/9706006.
\end{references}
\end{document}